\documentclass[%
 reprint,
showpacs,preprintnumbers,
 amsmath,amssymb,
 aps,prl, 
longbibliography, 
floatfix,
]{revtex4-1}

\usepackage{graphicx}
\usepackage{epstopdf}
\usepackage{dcolumn}
\usepackage{bm}
\usepackage{hyperref}
\usepackage[normalem]{ulem}

\newcommand{\ie}{\textit{i.e.}, }

\newcommand{\md}{\mathrm{d} }
\newcommand{\tbf}{\mathrm{bf} }
\newcommand{\tbb}{\mathrm{bb} }

\begin{document}


\title{Plasma Wave Seed for Raman Amplifiers}
\author{Kenan Qu}
\author{Ido Barth}%
\author{Nathaniel J. Fisch}%
\affiliation{Department of Astrophysical Sciences, Princeton University, Princeton, New Jersey 08544, USA }%

\date{\today}

\begin{abstract}
It is proposed to replace the traditional counterpropagating laser seed in backward Raman amplifiers with a plasma wave seed. 
In the linear regime, namely, for a constant pump amplitude,  a plasma wave seed may be found by construction that strictly produces the same output pulse as does a counterpropagating laser  seed.
In the nonlinear regime, or pump-depletion regime, the plasma-wave-initiated output pulse can be shown numerically to approach the same self-similar attractor solution for the corresponding laser seed.  
In addition, chirping the plasma wave wavelength can produce the same beneficial effects as chirping the seed wave frequency.
This methodology is attractive because it avoids issues in preparing and synchronizing a frequency-shifted laser seed.

\end{abstract}

\maketitle


To overcome the thermal damage limit of conventional materials~\cite{CPA1985, OPCPA1992}, it has been proposed to employ plasma to mediate intense laser amplification~\cite{PRL-Shvets1998, PRL-Malkin1999}.
Different backscattering coupling techniques were then explored, including Compton scattering~\cite{PRL-Shvets1998, ExpPRL2004-SRA}, resonant Raman scattering~\cite{PRL-Malkin1999, ExpPRL2004-Ping, ExpPRL2005-Suckewer, Ren_08,*ExpNat2007-Suckewer, yampolsky11,EXPPRL2008-Pai, EXPPoP2009-Ping, PoP-2014SRS}, Brillouin scattering~\cite{PoP-SBS-2006, EXPPRL2010-SBS, PRL-2013SBSExp, PRL-2016SBSExp, PRL2016-Matthew}, magnetized scattering~\cite{PRE-Yuan2017}, and chirped pump Raman amplification~\cite{PRL-Ersfeld2005}. 
In a Raman amplifier, a long pump laser pulse deposits its energy to a counterpropagating short amplified pulse, mediated by a plasma wave. 
The amplified pulse duration contracts while its amplitude grows, thereby producing an ultraintense laser pulse. 
To initiate the amplification, the amplified pulse is seeded by a short laser seed, down-shifted from the pump frequency by the plasma frequency,  and synchronized to meet the pump as the pump leaves the plasma. Preparing such a frequency-shifted, synchronized laser seed represents a significant technological challenge.

\begin{figure}[b]
\centering
\includegraphics[width=0.4\paperwidth]{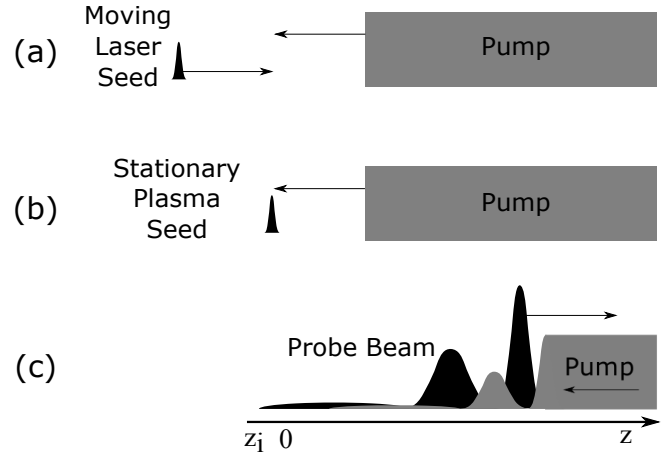}
\caption{ Seeding of the amplified output pulse. The pump beam is injected at the right. (a)  Triggered by a counterpropagating laser seed.   (b) Triggered by a  stationary plasma wave seed.  (c)  In the case of an advantageous result, the amplified output pulse assumes a $\pi$-pulse shape accompanied by depletion of the pump laser. }
\label{schm}
\end{figure}

It has  been proposed for strongly coupled Brillouin scattering to form  a laser seed by reflection of the  pump  \cite{peng2016single}.  
However, a simple reflection does not produce the frequency down-shift necessary for Raman resonance.
It is proposed here instead to replace the laser seed with a plasma wave seed. 
The plasma wave  has a negligible group velocity, so without regard for synchronization,  the amplification process is triggered only when the pump wave reaches it.   
In Fig.~\ref{schm}, we compare   seeding by a laser pulse (a) and by a plasma wave (b). 
For  laser seeding,  there exist solutions in which an intense counterpropagating wave is produced, with pump depletion (c), in which the counterpropagating pulse consumes essentially all the energy of the pump beam, while  contracting in time \cite{PRL-Malkin1999}.   
These solutions give rise to extreme intensities.  
However, there are other solutions, depending on how the laser seed is prepared with a front that is not steep enough, or with not enough initial energy content, in which these promising solutions do not obtain \cite{PRL2002-Tsidulko}.
The question to be answered here  is whether one can construct with a plasma wave seed the same promising solutions that were constructed with a suitably constructed laser seed.   

What is shown here is that, in fact, such solutions for the counterpropagating pulse, which we shall call the ``probe pulse'',  can be constructed for the plasma wave seed as well, and that they are attractor solutions.   
For nearly all laser-seeded interactions resulting in complete pump depletion,  a plasma-wave-seeded interaction can be found to produce the same time-asymptotic output pulse.  
This reachable subset includes the  advantageous output pulses depicted in   Fig.~\ref{schm}(c).  
The key plasma parameters  for efficient amplification~\cite{malkin2014review} will remain the same. 
Moreover, we show that the advantages that accrue to  laser seeding through frequency chirping \cite{PRL2012-Toroker} can be realized as well  through plasma wave seeding with wavelength chirping.

To proceed, consider the   three-wave resonant Raman coupling process~\cite{kruer1988physics}.  
For the case of the seeded plasma wave,  an electron density ripple, $\delta n$, is produced with a wave vector $\vec{k}_f$. 
This ripple is produced in only a small localized region at the far end of the plasma, at the end where the pump wave exits. 
When encountered by the incident pump laser with a wave vector $\vec{k}_a$,  electrons oscillate at velocity $\vec{v}$, and hence induce a transverse current, $\vec{j} = -e\vec{v} \delta n$, with a wave vector $\vec{k}_b = \vec{k}_a - \vec{k}_f$, where $e$ is the elementary  charge. 
If the wave vector and frequency of the transverse current $\vec{j}$ are properly matched, a probe beam is generated. 
Importantly, for efficient Raman amplification,  the dispersion relation determines the choice of $\vec{k}_f$ such that $(\omega_a-\omega_f)^2 = \omega_p^2 + c^2(\vec{k}_a-\vec{k}_f)^2$, where $\omega_a$ is the pump frequency and $\omega_f = \sqrt{\omega_p^2 + 3v_\mathrm{th}^2 k_f^2}$ is the plasma seed frequency. 
Here, $\omega_p$ and $v_\mathrm{th}$ are the plasma frequency and the electron thermal velocity, respectively.  
For cold and underdense plasmas, $\omega_f\approx\omega_p \ll \omega_a$; hence, $k_f\approx 2 k_a$. 
Note that in the case of the laser seed, the frequency $\omega_b$ is red detuned from the pump frequency by $\omega_f$. 
In contrast,  the plasma wave seed should have the appropriate wavelength, and then the appropriately  red-tuned frequency probe is automatically generated.

To develop  an equivalence rule between laser  and plasma seeds, we make the reasonable assumption that similar linear solutions (constant pump amplitude) will transition to the same nonlinear asymptotic attractor solutions which feature pump depletion.  
This assumption, of course, must be checked.  
However, the linearization  allows us to rigorously construct  a plasma wave seed  with  Green's function response identical to that of a laser seed  that has the  desired wavefront sharpness and intensity. 

Consider the configuration as illustrated in Fig.~\ref{schm}(b). 
At $t=0$, the pump coming from the right edge meets the Langmuir wave at $z=0$. 
We describe the envelopes  of pump and probe lasers using the wave vector potentials $a$ and $b$, 
normalized such that the pump intensity is $I_a = 2c\varepsilon_0(\pi m_e c^2|a|/e\lambda_a)^2 = 1.37\times 10^{18} (a/\lambda_a[\mathrm{\mu m}])^2 [\mathrm{W\,cm}^{-2}]$, 
and similarly for $I_b$, $f$ is the envelope of Langmuir wave normalized to $V f = (e/m_ec)|E_e|$, 
where $V\approx \sqrt{\omega_a\omega_p}/2$ for underdense plasmas (\ie $\omega_a\approx \omega_b \gg \omega_p$ with $\omega_b$ being probe wave frequency) and linearly polarized optical beams. 
Without losing generality, we assume real $a$ and keep $b$ and $f$ complex. 
Here, $m_e$ is the electron mass, $E_e$ is the electrostatic field of the Langmuir wave, $c$ is the speed of light, and $\varepsilon_0$ is the permittivity of free space. 
For simplicity, we neglect, at this stage, the group velocity dispersion (GVD), relativistic nonlinearity, and kinetic effects.
The resonant three-wave equations in cold plasma can then be written as~\cite{PRL-Malkin1999} 
\begin{equation}\label{1}
a_t - ca_z = -V bf, \quad b_t + cb_z = V af^*,\quad f_t = V ab^*, 
\end{equation}
where the subscriptions $t$ and $z$ denote the partial time and $z$ derivatives, respectively.

In the linear stage, before the pump beam gets depleted (\ie $a$ remains constant), the solution to Eqs.~(\ref{1}) can be obtained~\cite{supp} for $b(t,z) = b_\text{f}(t,z) + b_\text{b}(t,z)$,
where 
\begin{align}
	b_\text{f}(t,z) &= \int \md z' G_\tbf(t,z-z') f_0^*(z'), \label{31} \\
	b_\text{b}(t,z) &= \int \md z' G_\tbb(t,z-z') b_0(z'), \label{32} 
\end{align}
are the components of the probe generated by a plasma seed $f_0(z)$ and a laser seed $b_0(z)$, respectively, and 
\begin{align}
	G_\tbf(t,z) &= \frac{\gamma_0}{c} I_0 (\xi) \cdot \Theta(t-\frac{z}{c}), \label{greenf} \\
	G_\tbb(t,z) &= \frac{1}{\gamma_0} \frac{\partial}{\partial t} G_\tbf(t,z) \nonumber \\
	&\!\!\!\!= \frac{\gamma_0}{c} \sqrt{\frac{z}{ct-z}} I_1 (\xi) \cdot \Theta(t-\frac{z}{c}) + \frac1c \delta (t-\frac{z}{c}), \label{greenb}
\end{align}
are the Green's functions associated with the corresponding seeds~\cite{SHVETS1997}. Here $I_0(\cdot)$ is the zeroth order modified Bessel function,  $\Theta(\cdot)$ is the Heaviside function, $\gamma_0=a_0 V$ denotes the linear temporal growth rate,  and $\xi = 2(\gamma_0/c)\sqrt{z (ct-z)}$. 

In  laser-seeded amplification, $f_0=0$ and $b_0\neq 0$. 
The probe beam comprises the initial laser seed  [the second term in Eq.~(\ref{greenb})] and the generated components from the three-wave process [the first term in Eq.~(\ref{greenb})]. The generated component grows quasiexponentially and hence eventually dominates the probe. For a given laser seed $b_0(z)$, we now look for a plasma seed $f_0(z)$ such that the resulting probe (in the linear regime) is the same, \ie $b_\text{f}(t,z)=b_\text{b}(t,z)$.

First, in analyzing the Green's functions, note that 
\begin{equation}\label{3}
	\frac{c}{\gamma_0} \frac{\partial}{\partial z} G_\tbf(t,z) 
 =  -G_\tbb(t,z) 	+ \frac{\gamma_0}{c} \sqrt{\frac{ct-z}{z}} I_1 (\xi) \cdot \Theta(t-\frac{z}{c}), 
\end{equation}
where the last term can be neglected at the wavefront $z-ct\ll z$.  
The identity (\ref{3}) suggests  taking the envelope of plasma wave seed to be
\begin{equation}\label{trans}
	f_0(z)= \frac{c}{\gamma_0} \frac{\partial}{\partial z}b_0^*(z), 
\end{equation}
so that the generated probe beam can be found using integration by parts
\begin{align}\label{4}
	b_\text{f}(t,z) &= \int \md z' G_\tbf(t,z-z') \frac{c}{\gamma_0} \frac{\partial}{\partial z'} b_0(z') \nonumber \\
	&= - \int \md z' b_0(z') \frac{c}{\gamma_0} \frac{\partial}{\partial z'} G_\tbf(t,z-z') \nonumber \\
	&\cong  \int \md z' G_\tbb(t,z-z') b_0(z').
\end{align}
Comparing Eqs.~(\ref{32}) and (\ref{4}) shows that the plasma wave seed, initiated according to Eq.~(\ref{trans}),  generates asymptotically a probe beam that has the identical wavefront which is generated by use of a laser seed. 

As the probe pulse propagates and grows, if initially short enough, it should eventually deplete the pump and enter into the nonlinear or so-called ``$\pi$-pulse" regime, characterized by the self-similar contracting pulse envelope \cite{PRL-Malkin1999}.  
The asymptotic equivalence expressed by Eq.~(\ref{trans})  suggests that asymptotically identical pulses in the linear regime should evolve to identical pulses in the nonlinear regime. 
In the limit of short  (delta function) laser seeds, and  correspondingly short plasma seeds, a formal equivalence can be made  \cite{supp}.
While the finite width case does not follow rigorously, we  show numerically that the equivalence, in fact, does extend to finite pulse widths.

\begin{figure}[t]
\centering
\includegraphics[width=0.4\paperwidth]{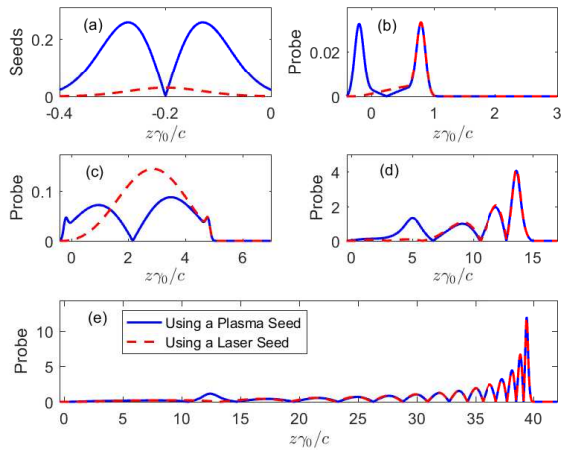}
\caption{(a) Envelopes of  plasma seed (blue solid line) and laser seed (red dashed line). (b)-(e) Amplified probe pulse at  times $\gamma_0t=1, 5, 15$, and $40$. Blue solid lines correspond to plasma seed;  red dashed lines to laser seed.  \label{fig1}}
\end{figure}

Thus, for finite width seeds, by numerically solving Eqs.~(\ref{1}), we compare amplification triggered by a plasma seed to that by a laser seed.
As shown in Fig.~\ref{fig1}(a), the laser seed envelope is  Gaussian, with $b_0(z) = 0.03 a_0 \exp[-(1+\gamma_0z/c)^2/\sigma^2]$, with normalized width $\sigma = 0.5$.
Following the equivalence rule in Eq.~(\ref{trans}), the plasma seed is taken as $f_0(z) = (c/\gamma_0)\partial b_0(z)/\partial z$. 
The two peaks in $f_0$ differ by a phase difference of $\pi$.  In Fig.~\ref{fig1}(b), we observe,  at $\gamma_0t=1$, that the probe beam generated by the plasma seed has a wavefront that exactly matches the laser seed.  The identical wavefronts generated by different seeds ensure the same Raman amplification throughout  $\gamma_0t=40$, as shown in Figs.~\ref{fig1}(b)-(e).  They both reveal similar $\pi$-pulse structures, with the amplitude in its leading peak increasing linearly and the pulse duration similarly contracting linearly in distance.

While we have constructed an  envelope of the plasma seed that approaches the corresponding attractor solution of the laser seed, it is not necessarily a unique correspondence.   Note that, while the laser seed has but one,  the plasma seed in Fig.~\ref{fig1}(a) has two maxima or two humps.   However, the asymptotic solution clearly depends on only the first hump since the pump interaction with that hump will clearly shadow the second hump. Indeed, removing the second hump will give the identical asymptotic solution, which, in practice, makes easier the setting up of the plasma wave seed.

\begin{figure}[b]
\centering
\includegraphics[width=0.4\paperwidth]{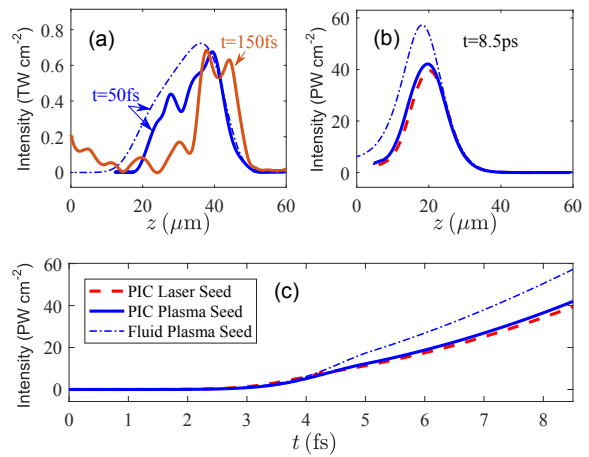}
\caption{Amplified probe pulses using a plasma seed (blue solid line) and a laser seed (red dashed line) at interaction time $t=50\mathrm{fs}$ (a) and $t=8.5\mathrm{ps}$ (b). (c) Comparison of the amplified probe pulse peak intensity. The thick solid and dashed lines are PIC simulations; and the thin dotted-dash lines are  fluid-model simulations using Eqs.(\ref{1}).  \label{fig4}}
\end{figure}

To examine kinetic effects using laser and plasma seeds, we conducted  one-dimensional particle-in-cell (PIC) simulations (using the code EPOCH \cite{EPOCH2015}).
For definiteness, we considered parameters  similar to those of recent experiments~\cite{EXPPoP2009-Ping, ExpPRL2005-Suckewer}. 
The electron density is $n_e=1.5\times 10^{19}\mathrm{cm}^{-3}$ and accordingly $\omega_p=2\pi\times35 \,\mathrm{THz}$. 
The pump laser has wavelength $\lambda_a=0.8\,\mu\mathrm{m}$ and intensity $I_a=0.4\,\mathrm{PW\,cm}^{-2}$. 
The linear growth rate is $\gamma_0=5\, \mathrm{ps}^{-1}$. Electrons are initialized to $10\,\mathrm{ eV}$, which avoids Landau damping~\cite{Supp3}.  
A cell size of $4\,\mathrm{ nm}$ is used to match the Debye length, with  $800$ electrons per cell  to reduce  charge density fluctuations.  
Calculations are done  in a $60\,\mu\mathrm{m}$ window moving with the group velocity  of the probe  ($v_g=0.9948c$). Collisions and ion  motion are ignored. 

The simulation results are shown  in Fig.~\ref{fig4}. 
The simulation includes, in addition to kinetic effects,   effects such as GVD and  relativistic nonlinearity, neglected in the hydrodynamic analysis. The laser seed  is Gaussian  with width $10\,\mu\mathrm{m} \approx 0.16c/\gamma_0$, and  peak intensity  $0.8\, \mathrm{TW\,cm}^{-2}$.  Its frequency is set at $\omega_b=2\pi\times340\, \mathrm{THz}$, obeying the frequency matching. For the plasma wave seed, as discussed, the  double peak structure with a $\pi$-phase difference may be eschewed in favor of a single peak. Thus, we use a Gaussian  to approximate the leading peak, with a normalized width $7.1\,\mu\mathrm{m} \approx0.117\,c/\gamma_0$, while  ignoring the second peak. The amplitude of the electrostatic seed wave is $5\times 10^9\, \mathrm{V/m}$, which is associated with a  $50\%$  electron density oscillation. Its wavelength is  $0.4\, \mu\mathrm{m}$, so that $k_p=2k_a$.

Note that, immediately after the interaction, a probe beam is generated [see Fig.~\ref{fig4}(a)]. 
For comparison, we also show the numerical solution of Eqs.~(\ref{1}). 
The comparison indicates that the plasma seed indeed triggers the Raman amplification like a laser seed does. 
No precursors are observed, but the moving window would suppress those. 
In Fig.~\ref{fig4}(c), which shows the growth of the peak intensity, we  identify the linear stage amplification (before $\sim 4.5\mathrm{ps}$),  exhibiting an intensity exponentially increasing  in time (or distance)  and the nonlinear stage (after $\sim 4.5\mathrm{ps}$), exhibiting  a  quadratically increasing intensity. 
After an amplification time of $8.5\,\mathrm{ ps}$, the leading spikes of the probes shown in Fig.~\ref{fig4}(b) both reach $40\, \mathrm{PW\,cm}^{-2}$, which is $100$ times higher than the pump intensity. 
The pump depletion at the probe peak is $75\%$ for both seeds. 
The agreement between the PIC simulations with plasma and laser seeds is very good for both the leading peak envelopes [Fig.~\ref{fig4}(b)] and the maximum intensity [Fig.~\ref{fig4}(c)]. 
They also decently match the numerical solutions of Eqs.~(\ref{1}), although PIC simulations show asymptotically lower peak intensities. 
The discrepancy might be due to kinetic effects that are taken into account in  PIC, but not in the fluid model, or due to the envelope approximation in the fluid model.


For higher plasma density  where GVD becomes important, one can advantageously chirp the laser seed  to reduce the required plasma length~\cite{PRL2012-Toroker}. 
Since higher frequency components propagate faster, the front of the chirped probe sharpens due to GVD. 
It is of interest to inquire whether  the same advantages can accrue for a plasma wave seed.
In fact, for plasma seeds, the advantageous chirping effects  can be accomplished by chirping the wavelength of the plasma wave. When it scatters off the monochromatic pump, the generated probe is also chirped and hence can contract due to GVD. 

\begin{figure}[b]
	\centering
	\includegraphics[width=0.42\paperwidth]{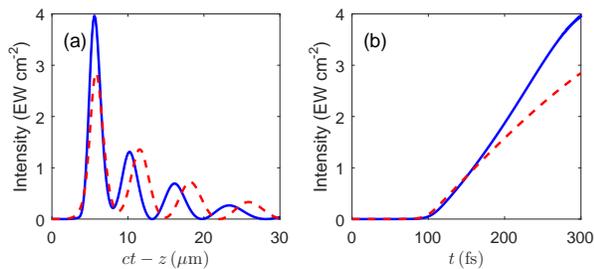}
	\caption{(a) Amplified probe
		beams triggered by chirped (blue solid) and  nonchirped (red dashed) plasma seeds with a $600 \, \mathrm{ fs}$ pump pulse;  (b)  growth of the peak intensities. The seed amplitudes and shapes are identical. }
	\label{figchirp}
\end{figure}

To show this,  we modify  Eqs.~(\ref{1}) by including the GVD term  for the probe~\cite{PRL2012-Toroker} 
$  b_t + c_b b_z = V af^* + i\kappa b_{tt}$,
where $\kappa = (1/2c_b) (\partial c_b/\partial\omega_b) = \omega_p^2/(2c_b \omega_b)$, and $c_b$ is the group velocity of the seed. 
In Fig.~\ref{figchirp}, we compare amplification in a high-density plasma with a chirped plasma seed to amplification with  a nonchirped plasma seed.
The parameters are chosen similar to those in Ref.~\cite{PRL2012-Toroker}, \ie $n_e = 12\times10^{20} \mathrm{cm}^{-3}$, $\lambda_a = 0.351\, \mu\mathrm{m}$, $I_a = 12.2 \,\mathrm{PW\,cm}^{-2}$, and the plasma length is $90\,\mu\mathrm{m}$. 
The plasma seeds are both Gaussian  with  FWHM of $1.8\,\mu\mathrm{m}$. 
The nonchirped seed has a uniform wave vector $k_f = k_a+k_b = 2\pi/0.215\, \mu\mathrm{m}$. 
Its output intensity reaches $2.8\, \mathrm{EW\,cm}^{-2}$. 
Here, the wave vector, $k_f$, of the chirped plasma seed increased $3.5\%$ per $\mu\mathrm{m}$, where pump interacts with lower $k_f$ first. Since the pump has a constant frequency, the generated probe is also chirped with smaller wave numbers (smaller frequencies) at the front. Because of GVD, the probe contracts when traveling through the plasma.
Similar to the case of a chirped laser seed \cite{PRL2012-Toroker}, the probe beam has a larger growth rate and its intensity reaches $4\, \mathrm{EW\,cm}^{-2}$. From Fig.~\ref{figchirp}(a), we also observe an appropriately   narrower probe pulse. 

Can any laser seed be replaced by an equivalent plasma wave seed?  
Clearly, for a seed laser pulse envelope that is not differentiable, the formal equivalent plasma seed envelope, according to Eq.~(\ref{trans}), does not exist.
However, this restriction on shape is not important since the seed laser pulse need be no more sharp than a Gaussian in order to access the $\pi$-pulse pump depletion regime  \cite{PRL2002-Tsidulko}, for which  equivalent plasma seeds do exist.   
More relevant is the  restriction on the amplitude; while a seed laser pulse envelope can take an arbitrarily large amplitude, the Langmuir wave seed amplitude will be limited by  the wave breaking limit.  
In a cold plasma, this condition is equivalent  to a maximum density variation, $\delta n < n_0$;
in a warm plasma, the density variation is somewhat more limited \cite{coffey1971breaking, LangWaveLimit1988}.
However, this will not be an issue in the main amplification regimes~\cite{Clark_03}, for which the necessary laser seeding amplitudes can be small \cite{Yampolsky2004PRE}.
Also, even a small plasma wave seed can access operation near the wave breaking limit \cite{PoP2015-Matthew, PoP2014-Toroker}, since that limit is determined by the pump amplitude, not the seed amplitude.
The density restriction may be an issue, though in the so-called {\it quasitransient} regime, where damping of the Langmuir wave is significant, so amplification is achieved only with larger laser seed amplitudes  \cite{Malkin2009PRE}.  
The corresponding amplitudes may then not be available for plasma wave seeds.
However, these regimes are, in any event, not of interest for plasma wave seeding since, if the plasma wave is heavily damped, the advantage of synchronization is absent.

In summary, a Langmuir wave seed can, in principle,  replace almost any useful laser seed in backward Raman amplifiers,  generating a probe beam that has an identical wavefront. 
The method can also be generalized by using a chirped plasma wave wavelength to mimic the frequency chirping of a laser seed.
These  predictions  are supported by hydrodynamic and fully kinetic 1D-PIC simulations.  
The promise of this method is that it represents an alternative technology in implementing compression of high-intensity lasers in plasma, with a particular advantage concerning the timing of the pulses.

Although the technological production of the plasma wave seed is beyond the scope of this work, one straightforward methodology would be to employ low-power high-quality counterpropagating laser pulses to produce the plasma wave seed; this seed would then linger in the plasma (it has near zero group velocity) until the high-	power pump laser (which needs not be of high quality) excites the parametric interaction.  In a plasma setting, localized plasma waves have been generated by stimulated Raman scattering using a tightly focused intense laser pulse in preformed plasma~\cite{EXP-PRL2005-pseed, EXP-PRL2006-pseed, EXP-PRL2009-pseed, EXP-PRL2016-pseed}.  The ability of plasma wave seeds to linger in plasma and then to scatter laser energy has already been exploited in a variety of settings, including plasma holography~\cite{PRL2001-Dodin}, plasma gratings~\cite{PlasmaGratings2014} and plasma photonic crystals~\cite{Lehmann2016, Lehmann2017}.  In dynamic Brillouin  gratings in optical fibers and photonic chips~\cite{DongDBG2015, SantagiustinaDBG2013}, the acoustic mode, playing the role of the plasma wave, can be arranged similarly to retain information.  While the plasma seed is lingering, it might be manipulated to better serve as a seed, for example, through autoresonant techniques~\cite{PRL-Barth2015}. The possibilities outlined here should serve to stimulate the optimizing of means for generating the plasma wave seed.  Moreover, the methodology offered here can be generalized to other waves with negligible group velocity that might mediate high-intensity laser compression in plasma, such as, for example, the ion acoustic wave for compression by stimulated Brillouin scattering.

\begin{acknowledgments}
This work was supported by NNSA Grant No. DE-NA0002948, and AFOSR Grant No. FA9550-15-1-0391. 
\end{acknowledgments}

\bibliography{Plasma-ref}

\end{document}